\documentclass[prc,twocolumn,superscriptaddress,showpacs]{revtex4}
\usepackage{graphicx}
\usepackage{amsmath,amssymb}
\usepackage{mathrsfs}
\usepackage{verbatim}
\usepackage{epic,eepic}
\usepackage{float}
\usepackage{color}
\usepackage[colorlinks = true,linkcolor = blue,citecolor = blue,urlcolor = blue,filecolor = blue,pagecolor = blue, breaklinks=true]{hyperref}

\begin{document}

\title{Competition between fermions and bosons in nuclear matter at low densities and finite temperatures}

\author{J.~Mabiala}
\email[]{justin.mabiala@lnl.infn.it} 
\affiliation{INFN, Laboratori Nazionali di Legnaro, Italy}
%\altaffiliation{Present address: INFN, Laboratori Nazionali di Legnaro, Italy; justin.mabiala@lnl.infn.it}
%\affiliation{Cyclotron Institute, Texas A$\&$M University, College Station, Texas 77843, USA}
\author{H.~Zheng}
\email[]{zheng@lns.infn.it}
\affiliation{Laboratori Nazionali del Sud, INFN, via Santa Sofia, 62, 95123 Catania, Italy}
%\altaffiliation{Present address: Laboratori Nazionali del Sud, INFN, via Santa Sofia, 62, 95123 Catania, Italy}
%\affiliation{Cyclotron Institute, Texas A$\&$M University, College Station, Texas 77843, USA}
%\affiliation{Physics Department, Texas A$\&$M University, College Station, Texas 77843, USA}
\author{A.~Bonasera}
\affiliation{Laboratori Nazionali del Sud, INFN, via Santa Sofia, 62, 95123 Catania, Italy}
\affiliation{Cyclotron Institute, Texas A$\&$M University, College Station, Texas 77843, USA}
\author{Z.~Kohley}
\affiliation{National Superconducting Cyclotron Laboratory, Michigan State University, East Lansing, Michigan 48824, USA}
%\altaffiliation{Present address: National Superconducting Cyclotron Laboratory, Michigan State University, East Lansing, Michigan 48824, USA}
%\affiliation{Cyclotron Institute, Texas A$\&$M University, College Station, Texas 77843, USA}
%\affiliation{Chemistry Department, Texas A$\&$M University, College Station, Texas 77843, USA}
\author{S.~J.~Yennello}
\affiliation{Cyclotron Institute, Texas A$\&$M University, College Station, Texas 77843, USA}
\affiliation{Chemistry Department, Texas A$\&$M University, College Station, Texas 77843, USA}

\date{\today }

\begin{abstract}
%\textcolor{blue}{}
We derive the free energy for fermions and bosons from fragmentation data. Inspired by the symmetry and pairing energy of the Weizs\"acker mass formula we obtain the free energy of fermions (nucleons) and bosons (alphas and deuterons) using Landau's free energy approach. We confirm previously obtained results for fermions and show that the free energy for alpha particles is negative and very close to the free energy for ideal Bose gases. Deuterons behave more similarly to fermions (positive free energy) rather than bosons. This is due to their low binding energy, which makes them very `fragile', i.e., easily formed and destroyed. We show that the $\alpha$-particle fraction is dominant at all temperatures and densities explored in this work. This is consistent with their negative free energy, which favors clusterization of nuclear matter into $\alpha$-particles at subsaturation densities and finite temperatures. The role of finite open systems and Coulomb repulsion is addressed.
 
\end{abstract}

\maketitle

%\section{INTRODUCTION}

%\section{EXPERIMENTAL DETAILS AND EVENT SELECTION}

%\section{TEMPERATURE AND DENSITY}

%\section{RESULTS AND DISCUSSION}

Dilute mixed systems composed of fermions and bosons exhibit a large variety of interesting features that have been the subject of several theoretical and experimental works. Although generally considered as made of strongly interacting fermions (protons and neutrons), nucleonic systems have been observed to display some properties relevant of bosons. Some of these aspects are the $\alpha$-decay in heavy nuclei, preformed $\alpha$-particles in the ground state of nuclei \cite{PhysRevC.82.031301} and the cluster structure of $N$=$Z$=even light nuclei \cite{Freer20141}. While the tunneling through the Coulomb barrier is well understood, the preformation of the $\alpha$-particle is still a difficult task for theoretical model descriptions.

Experiments in heavy-ion reactions at energies around the Fermi energy have revealed the creation of dilute nuclear matter in which the strong interaction has led to the emergence of correlated states of nucleons (clusters). These few-body correlations remain substantial even at very small densities ($\backsim0.01\rho_0$ or less; $\rho_0$ = 0.15 nucl/fm$^{3}$) and at moderate temperatures \cite{PhysRevC.75.014601,PhysRevLett.104.202501,PhysRevC.85.064618,PhysRevLett.108.172701}. In fact, at low densities the system can minimize its energy by forming light clusters such as deuterons or strongly bound $\alpha$-particles. Clustering effects are expected to modify the density dependence of the symmetry energy of nuclear matter \cite{Horowitz200655,Eur.Phys.J.A50.17}, and the structure of atomic nuclei \cite{Freer20141}. 

The thermodynamic properties of nuclear matter play an important role in studies of various astrophysical phenomena \cite{Danielewicz22112002, Lattimer23042004, li2008recent, Giuliani2014116}. Knowledge of thermodynamic quantities such as the free energy of fragments is needed when considering a wide range of temperatures, densities and/or proton fractions. In fact, the free energy of fragments defines the balance between denser fragments and the more dilute nucleonic gas. Its changes with temperatures and densities are of crucial importance to better understand the properties of dense nuclear matter.

In this paper, we report on experimental free energy (density) for fermions and bosons from the fragmentation of quasiprojectiles by application of Landau's free-energy approach. The temperature and density of the produced quasiprojectile systems are determined using the quantum-fluctuation method, fully described in Refs.~\cite{Zheng:2010kg,PhysRevC.86.027602,Zheng201243,PhysRevC.88.024607,JPGNuclPartPhys.41.055109}. We notice, and it is an important result, that the free-energy density for alphas is negative. In contrast, it is positive for deuterons,  and close to that for fermions. The free-energy density for ideal Bose gases gives results similar to those for alphas but has opposite sign for those of deuterons. This demonstrates that alphas behave indeed as bosons while deuterons do not, which is due to their low binding energy, thus being continuously created and destroyed during the time evolution of the system. The fact that the free-energy density is negative means that if $N$=$Z$=even systems will `live' long enough, all the particles will cluster into alphas while deuterons will disappear.  

The experiment was performed at the Cyclotron Institute, Texas A$\&$M University. Beams at 35 MeV/A of $^{64}$Zn, $^{70}$Zn, and $^{64}$Ni from the K-500 superconducting cyclotron were used to respectively irradiate self supporting targets of $^{64}$Zn, $^{70}$Zn, and $^{64}$Ni. The 4$\pi$ NIMROD-ISiS setup \cite{Wuenschel2009578,Schmitt1995487} was used to collect charged particles and free neutrons produced in the reactions. A detailed description of the experiment can be found in Refs.~\cite{KohleyPhD,PhysRevC.83.044601, PhysRevC.86.044605}. For events in which all charged particles were isotopically identified, the quasiprojectile (QP) was reconstructed using the charged particles and free neutrons. This reconstruction includes, therefore, determination of the QP composition, both $A$ and $Z$. The neutron ball provided event-by-event experimental information on the free neutrons emitted during a reaction \cite{Wuenschel20101,Marini201380}. Particles with $Z$=1, 2 and $Z\geq$3, detected by  NIMROD-ISiS setup were attributed to QP decay when their longitudinal velocities lay within the range of $\pm65\%$, $\pm60\%$, $\pm40\%$, respectively, of the coincident QP residue velocity in the event \cite{PhysRevC.79.061602,Wuenschel20101}. Thermally equilibrated
QP events were selected by requiring the QP to be on average spherical, in a narrow range of shape deformation. The sum of the masses of the collected and accepted fragments was constrained to be in the range of $54\leq A_{QP} \leq 64$. Events were then sorted in 8 QP excitation energy bins, 1 MeV/A wide, ranging from 2.5 to 9.5 MeV/A.
 
%%%%%%%%%%%%
%%% Fig1  %%
%%%%%%%%%%%%
\begin{figure}[h]
\includegraphics[scale=0.4]{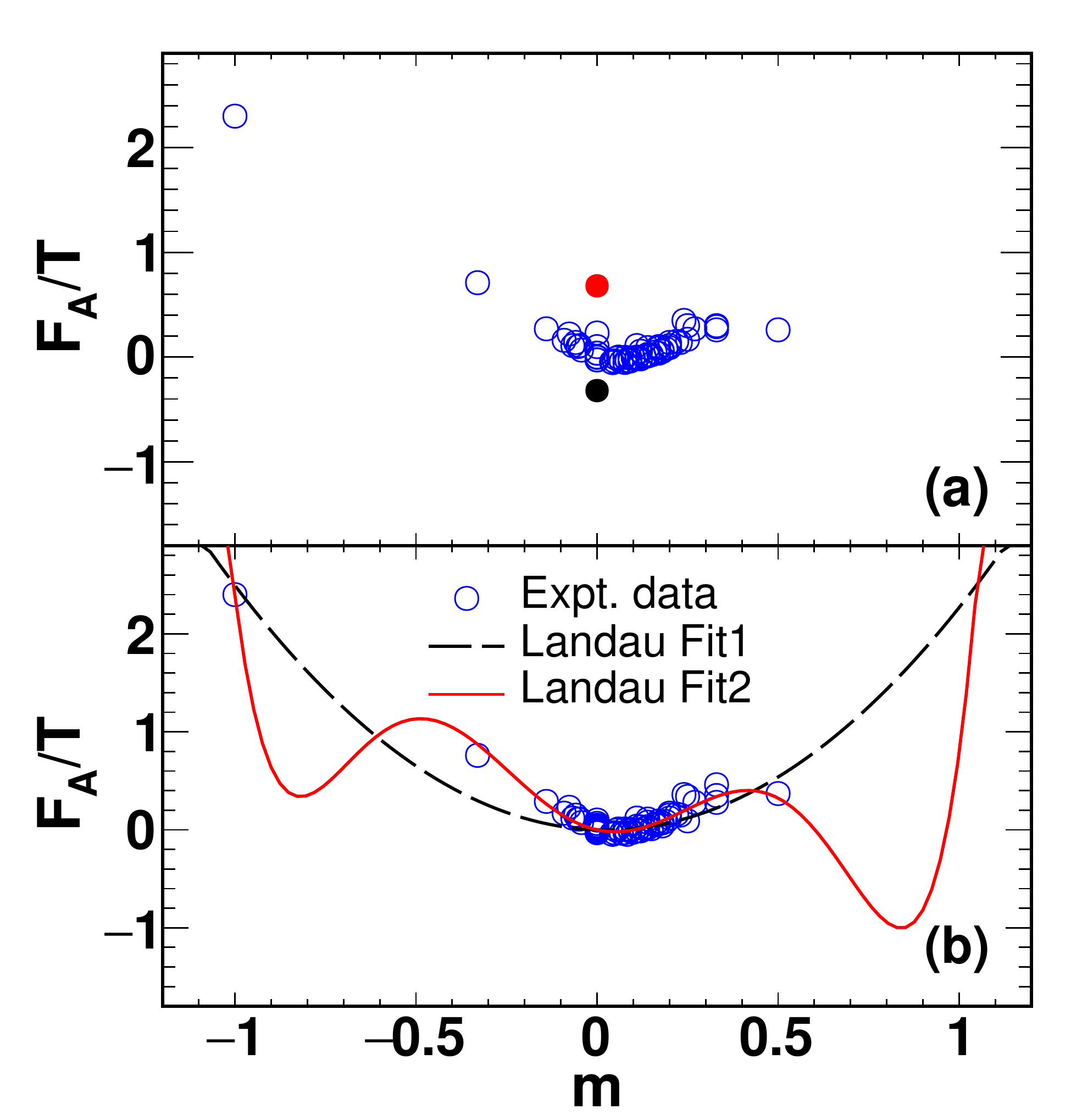}
\caption{(Color online) $F_A/T$ for fragments as a function of fragment's neutron-proton asymmetry $m$ for an excitation energy of 5.5 MeV/A of the QP. (a) $F_A/T$ values calculated from fragment yield data normalized to the yield of $^{12}$C. Data points corresponding to deuteron and $\alpha$ are colored in red and black, respectively. (b) $F_A/T$ values calculated after correcting for pairing effects using $a_p/T$ values obtained from the analysis of $N$=$Z$ nuclei \cite{PhysRevLett.101.122702,PhysRevC.81.044618,PhysRevC.83.054609,PhysRevC.87.017603}. The dashed line (Landau Fit1) is a fit to data using only the first and last terms of Eq.~(\ref{Eq1}). The solid line (Landau Fit2) represents a fit to data using the complete Landau free energy [Eq.~(\ref{Eq1})] with $a$, $b$, $H/T$ as free parameters, and fixing $c=115$ that was observed to be almost constant, within uncertainties, over the entire range of the QP excitation energy \cite{PhysRevC.87.017603}. The values of $a$, $b$, and $H/T$ corresponding to the solid line were obtained as 16.289$\pm$0.024, -102.871$\pm$0.058 and 0.808$\pm$0.002, respectively. Error bars corresponding to statistical errors are smaller than the symbols.}
\label{Fig1}
\end{figure}

Recently, we have analyzed fragment yield data to investigate the nuclear phase transition using the Landau free energy technique \cite{PhysRevLett.101.122702,PhysRevC.81.044618,PhysRevC.83.054609,PhysRevC.87.017603}. This approach is based on the assumption that, in the vicinity of the critical point, the fragment free energy per nucleon ($F_A$) relative to the system temperature ($T$) can be expanded in a power series in the fragment's neutron-proton asymmetry $m$ as 

\begin{equation}
\dfrac{F_A}{T}=\dfrac{1}{2}am^2+\dfrac{1}{4}bm^4+\dfrac{1}{6}cm^6-\dfrac{H}{T}m\ ,
\label{Eq1}
\end{equation}
where $m$=$(N-Z)/A$, and $N$, $Z$, and $A$ are the neutron, proton, and mass numbers of the fragment, respectively. The quantity $m$ behaves as an order parameter, $H$ is its conjugate variable and the coefficients $a$, $b$, and $c$ are fitting parameters. According to the modified Fisher model \cite{Fisher1967,Minich1982458,PhysRevLett.101.122702}, the fragment yield is given by $Y$=$y_0A_f^{-\tau}e^{-(F_A/T)A}$ near the critical point; with $\tau$=$2.3\pm0.1$ the critical exponent \cite{PhysRevLett.101.122702,Bonasera20001} and $y_0$ a constant.

Figure~\ref{Fig1}(a) shows the fragment free energy ($F_A/T$) values as a function of their neutron-proton asymmetry $m$ at an excitation
energy of 5.5 MeV/A of the QP. One clearly sees that $F_A/T$ values for $N$=$Z$ fragments ($m$=0) significantly deviate from the regular behavior of the $N$$\neq$$Z$ fragments. This shows the significant role of odd-even effects, which we will loosely refer to as pairing. We can generalize the Landau approach to include the free energy for $m$=0 particles, i.e., bosons. Inspired by the pairing energy per particle for a fragment with mass number $A$ ($E_p$=$a_p\delta/A^{3/2}$), one obtains the following linear equation for the analysis of $N$=$Z$ fragments
\begin{equation}
\ln (YA^{\tau})=\ln (y_0) +\frac{a_p}{T}\frac{\delta}{A^{1/2}}\ ,
\label{Eq2}
\end{equation}
where $a_p/T$ is the slope and $\ln(y_0)$ the intercept, and it is discussed in more details in Refs.~\cite{PhysRevC.83.054609,S021830131250019X,PhysRevC.87.017603}. The quantity $\delta$=-1, 0 and +1 for odd-odd, odd-even and even-even fragments, respectively; the term $\Pi$=$\delta/A^{3/2}$ to be the order parameter in Landau's description and plays the same role as the order parameter $m$ for the free energy. Higher order terms  in $\Pi$ might be possible but a good fit is obtained at this lowest order with the available data. A linear fit to the $N$=$Z$ data allows the extraction of the values of $a_p/T$ and $y_0$ \cite{PhysRevC.87.017603}.

In Fig.~\ref{Fig1}(b), $F_A/T$ values corrected by ``pairing" are shown. The dashed line (Landau Fit1) represents a fit to data using only the first and last terms of Eq.~(\ref{Eq1}), a case corresponding to a single phase as in the Weizs\"acker mass formula. The solid line (Landau Fit2) is a fit to data using the complete
Landau free energy as given by Eq.~(\ref{Eq1}). As the efficiency for measuring neutrons differs from the efficiency for measuring charged particles, neutron yields were excluded from the fitting. It is observed in the figure that the complete form of Eq.~(\ref{Eq1}) provides a better fit to the free energy data. The appearance of the three minima is a signature of a first-order phase transition of the system \cite{K.Huang,PhysRevLett.101.122702,PhysRevC.83.054609,PhysRevC.81.044618,PhysRevC.87.017603}.

%%%%%%%%%%%%
%%% Fig2  %%
%%%%%%%%%%%%
\begin{figure}[h]
\includegraphics[scale=0.4]{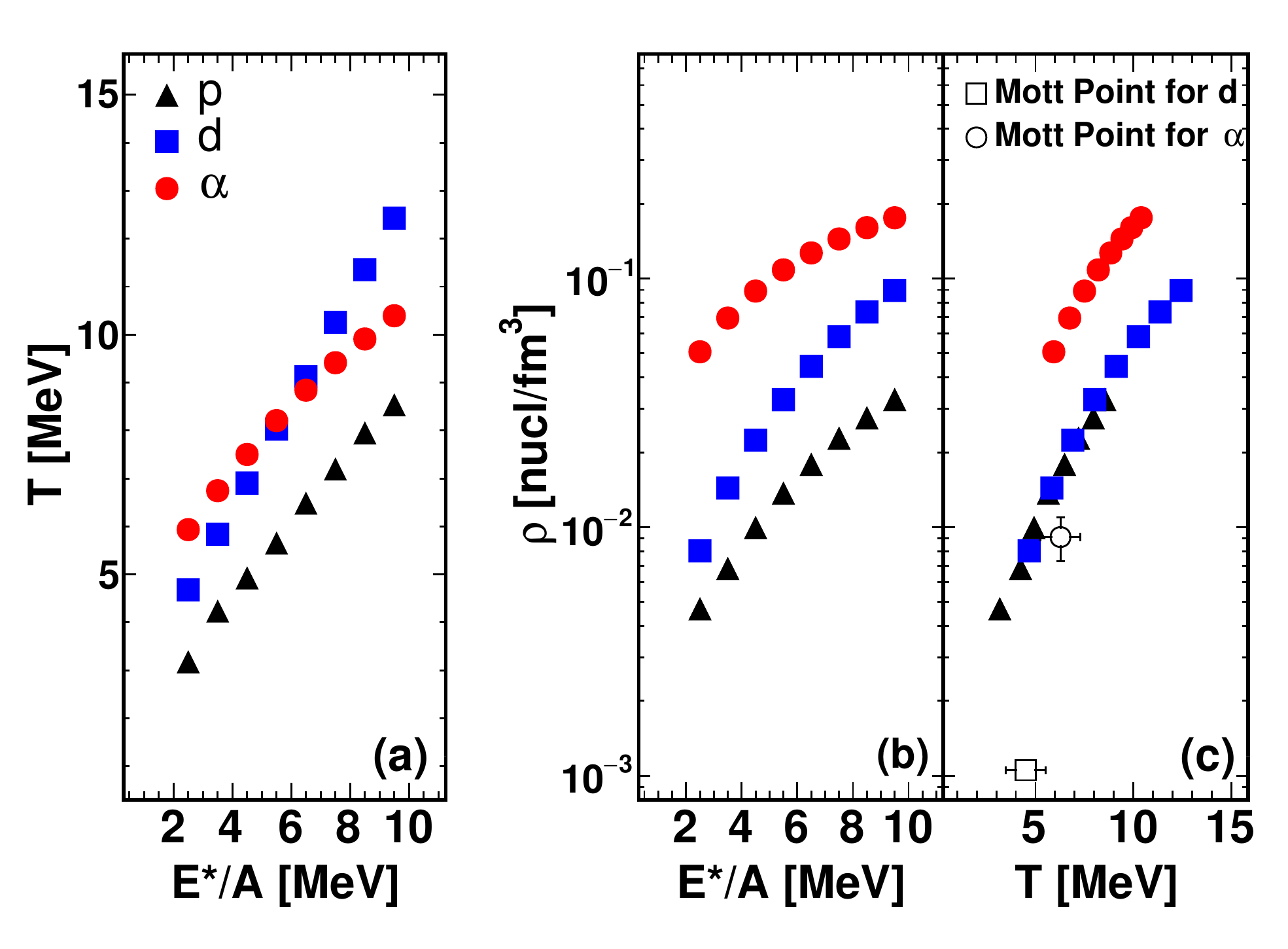}
\caption{(Color online) (a) and (b) Temperatures and densities sampled by three probe light particles ($p$, $d$ and $\alpha$) emitted by the QP as a function of its excitation energy. (c) Density plotted as a function of temperature for $p$, $d$ and $\alpha$. Error bars corresponding to statistical errors are smaller than the symbols. For comparison, experimentally derived Mott points for $d$ (empty square) and $\alpha$ (empty circle) are also shown~\cite{PhysRevLett.108.062702}.}
\label{Fig2}
\end{figure}

The temperatures and densities of the QP are determined from the fluctuations of the transverse momentum quadrupole $Q_{xy}$=$p^2_x-p^2_y$, average multiplicities and multiplicity fluctuations. These observables are used to correct for Coulomb effects as well. Further details can be found in Refs.~\cite{Zheng:2010kg,PhysRevC.86.027602,Zheng201243,PhysRevC.88.024607,JPGNuclPartPhys.41.055109}. We have applied the method of correcting for Coulomb effects to experimental data in Ref.~\cite{PhysRevC.90.027602}, and it was observed that the Coulomb corrections lower temperature values by almost 2 MeV, but the effects on the densities were observed to be small. A similar procedure was also applied to the data from INDRA-collaboration for $p$, $d$ and $\alpha$ \cite{Marini2016194} and their results are consistent with ours. In Ref.~\cite{PhysRevC.92.024605}, the error in applying the Coulomb corrections which arises from the uncertainty in the source charge was  estimated to be respectively $\pm$2$\%$ for the densities and $\pm$6$\%$ for the temperatures.

%%%%%%%%%%%%
%%% Fig3  %%
%%%%%%%%%%%%
\begin{figure}[h]
\includegraphics[width=8.5cm,height=6.cm,angle=0]{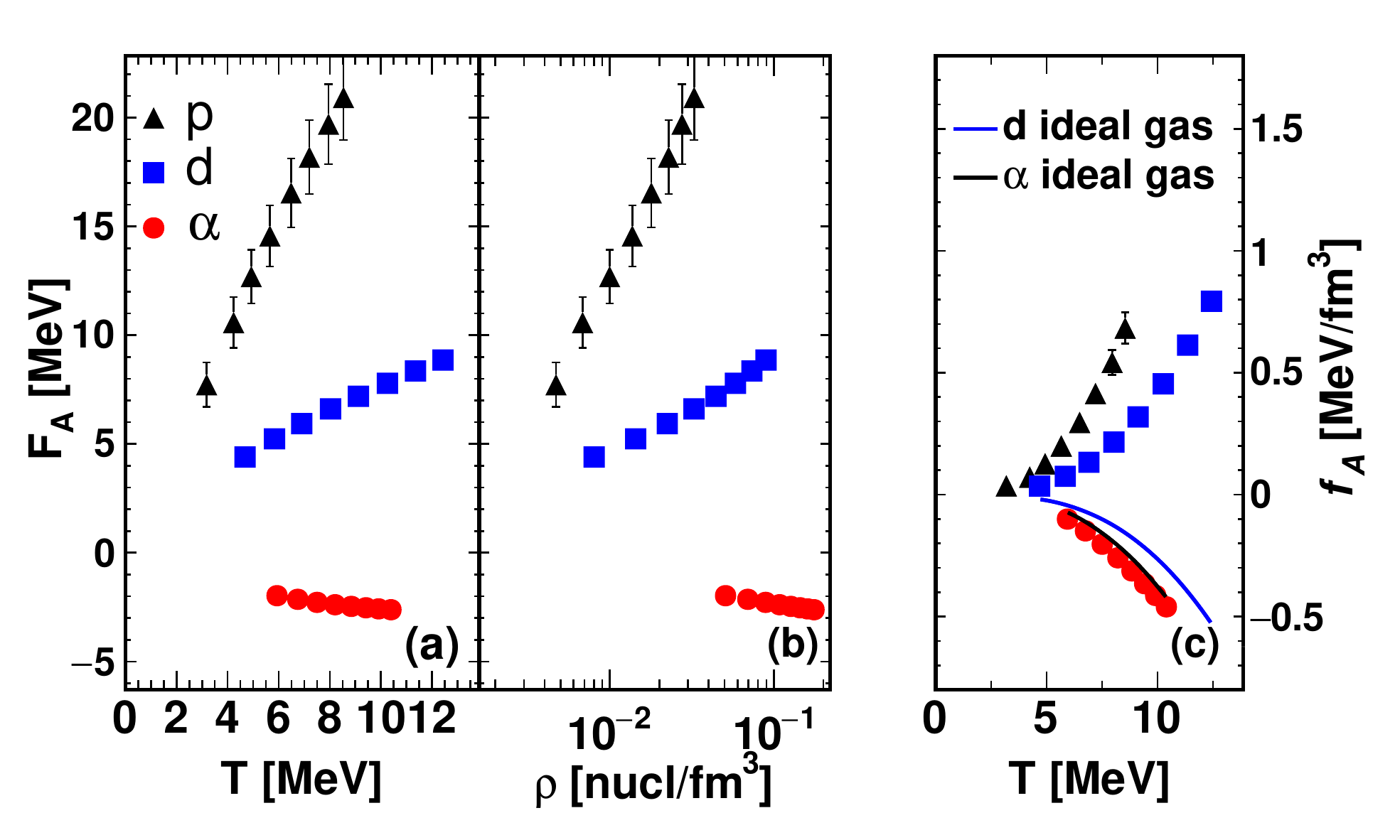}
\caption{(Color online) (a) and (b) Free energy for $p$, $d$ and $\alpha$ calculated within the framework of Landau's approach as a function of their sampled temperature and density. (c) Free-energy density ($f_A$=$F_A\times\rho$) as a function of temperature for the three light particles. Error bars are shown when statistical errors exceed the size of the symbols. Solid lines refer to the ideal gas free-energy density for bosons ($d$ and $\alpha$).}
\label{Fig3}
\end{figure}

The QP temperatures ($T$) and densities ($\rho$) sampled by protons ($p$), deuterons ($d$) and alphas ($\alpha$)  as a function of the excitation energy per nucleon of the reconstructed QP, $E$*/$A$, are shown in Fig.~\ref{Fig2}(a) and Fig.~\ref{Fig2}(b). Since the evaluation of associated errors on $T$ and $\rho$ was not straightforward, we have estimated the physical values from the difference in extracted values between the full and half datasets. Estimated statistical errors on $T$ and $\rho$ are smaller than the symbols (better than 3$\%$). These errors do not, however, include the uncertainty from Coulomb corrections. $T$ and $\rho$ values for the three light particles are observed to rise with $E$*/$A$. While $T$ values for $d$ and $\alpha$ are very close, the densities seen by the two particles are very different from each other. We also observe that densities probed by alphas are slightly higher than $\rho_0$ at highest $E$*/$A$ values, which we will discuss further later. We have to stress that densities and temperatures of bosons are derived under the assumption of Coulomb repulsion among fragments. Of course at higher densities these fragments will start to overlap and the attractive nuclear force might become dominant. Boson systems below the critical point (condensate) become unstable if an attractive force is at play \cite{K.Huang}, thus we expect our approximation to break down at some density and temperature. Since the temperature for which near ground-state densities are reached is rather high, the kinetic energy might be dominant with respect to the interactions and our approximation should still be valid. The correlation between the density and the temperature, as probed by $p$, $d$ and $\alpha$, is presented in Fig.~\ref{Fig2}(c). It is interesting to see that $p$ and $d$ display one single curve, even though there is a clear difference for the behavior of their sampled $T$ and $\rho$ with $E$*/$A$. The binding energy of a cluster relative to the medium vanishes at a point known as the Mott point\cite{PhysRevC.81.015803}. Since we observe alphas coming from high densities, we have shown the Mott points for $d$ and $\alpha$ obtained in Ref.~\cite{PhysRevLett.108.062702} for comparison. Note that the method to derive the Mott point \cite{PhysRevLett.108.062702} is based on classical approximations at variance with our quantum approach and we have used a different way to correct for Coulomb effects \cite{PhysRevC.90.027602}. 

After deriving $T$ values, we apply Eq.~(\ref{Eq1}), using the extracted Landau's fitting parameters, to determine the fragment free energy per nucleon, $F_A$, for fermions. For Bosons ($d$ and $\alpha$), we adopt the parametrization $F_A$=$-a_p\delta/A^{3/2}$ to easily derive the free energy. Figures \ref{Fig3}(a)-(b) depict the temperature and density dependence of the derived values of $F_A$ for $p$, $d$ and $\alpha$. Estimated statistical errors on $F_A$ for protons are 10$\%$ while those for $d$ and $\alpha$ are smaller than the symbols (better than 3$\%$). There is a strong correlation of increasing $F_A$ with increasing $T$ and $\rho$, for $p$ and $d$. In contrast to $p$ and $d$ results, $F_A$ values for $\alpha$ are negative and weakly depend on $T$ and $\rho$. From the values of $F_A$ and $\rho$, we examine in Fig.~\ref{Fig3}(c) the free energy density ($f_A$=$F_A\times\rho$) against $T$. It is observed that $f_A$ approaches zero in the limit $T\rightarrow0$ MeV, as expected, and differences between $p$ and $d$ curves seen in Fig.~\ref{Fig3}(a)-(b) are less pronounced. Figure \ref{Fig3}(c), also, shows $f_A$ results obtained for an ideal Bose gas (solid lines) \cite{Laundau.Lifshitz}. The positive experimentally-derived $f_A$ values for $d$ indicate that these particles behave much like fermions, i.e., they break and recombine easily. For a system in equilibrium, this implies that the system of nucleons will eventually coalesce into $\alpha$-particles. The obtained values of the free-energy density for alphas are below (in negative values) the ideal Bose gas values which confirms that the repulsive Coulomb force enhances condensation similar to atomic traps \cite{Marini2016194,PhysRevLett.87.080403,Zheng201243}.

%%%%%%%%%%%%
%%% Fig4  %%
%%%%%%%%%%%%
\begin{figure}[h]
\includegraphics[scale=0.4]{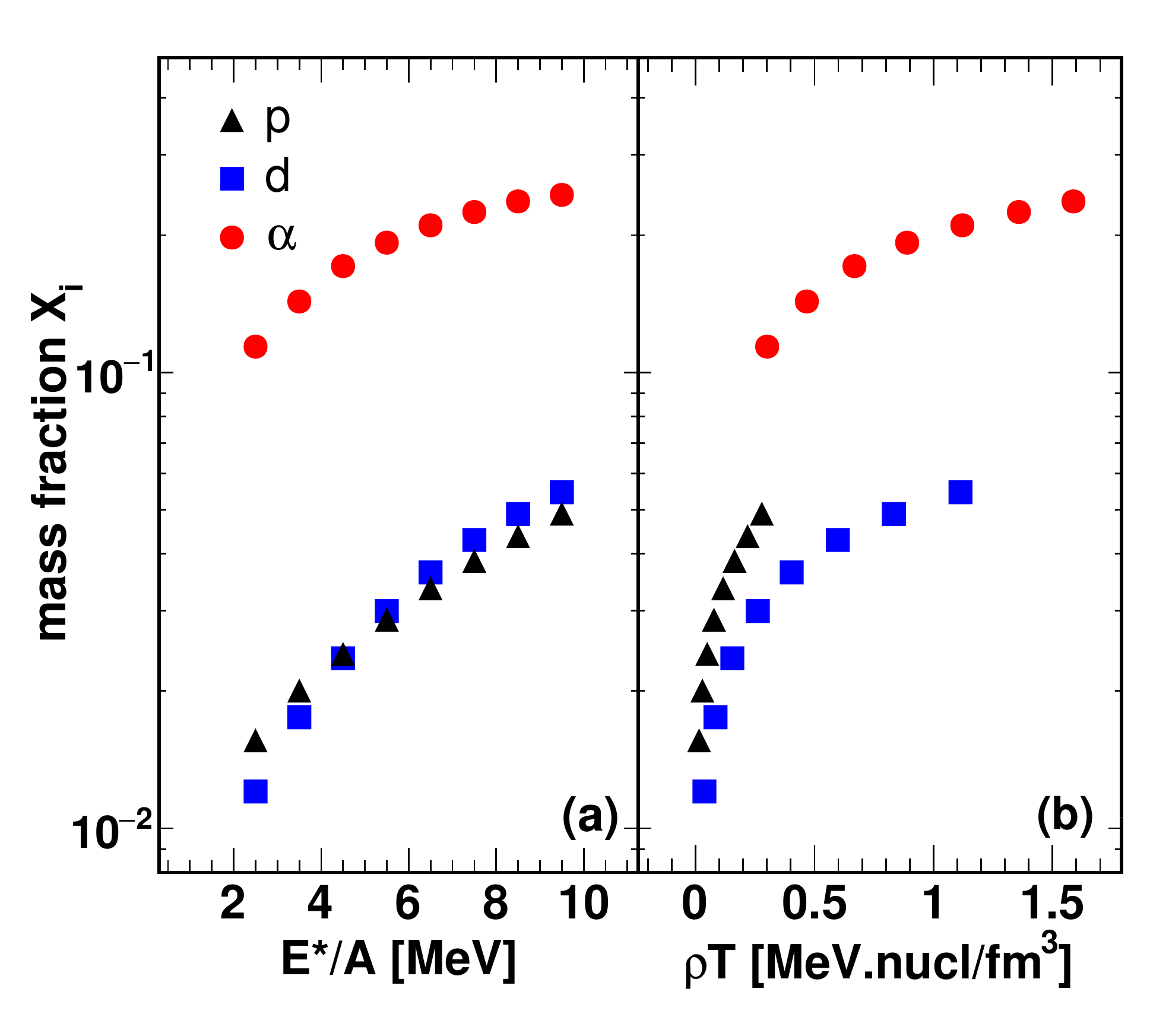}
\caption{(Color online) Mass fractions of $p$, $d$ and $\alpha$ are shown as a function of $E$*/$A$ (a) and of the `kinetic pressure' $\rho\times T$ (b). Statistical errors are smaller than the symbols.}
\label{Fig4}
\end{figure}

In theoretical models cluster mass fractions are commonly used to characterize the degree of clusterization in low-density matter. Figure \ref{Fig4}(a) shows mass fractions, $X_i$=$n_iA_i/A_{QP}$, of the three light particles as a function of $E$*/$A$, derived directly from data. The quantities $n_i$ and $A_i$ are, respectively, the multiplicity and mass of particle $i$, and $A_{QP}$ denotes the mass of the reconstructed fragmenting source ($A_{QP}\approx$60). While a higher $\alpha$-cluster fraction is seen for all $E$*/$A$ values, $p$ and $d$ have similar mass fractions. In Fig.~\ref{Fig4}(b), the behavior of $X_i$ is displayed as a function of the `kinetic pressure' $\rho\times T$. In Ref.~\cite{PhysRevC.81.015803}, in which a microscopic quantum statistical approach and a generalized relativistic mean-field model were employed, it was reported that complex particles may still appear beyond the Mott point, and $X_{\alpha}$ was found to decrease at high $\rho$. However, our results show that $X_{\alpha}$ is enhanced at high $E$*/$A$ and, correspondingly, at high $T$ and $\rho$. We believe this may be due to two main aspects. The first one is the fact that we have an open finite system. It could be that during the fast expansion, nucleons may still have the possibility to coalesce into $\alpha$-particles. Also we should mention the fact that in the theory of $\alpha$-decay, one assumes a preformation factor to explain the observations. This means that alphas are formed in the nuclear ground state, maybe at the surface. In any case, the large negative value of the derived free-energy density for alphas is consistent with the large mass fraction obtained directly from experimental data. We also mention the fact that for infinite systems Coulomb forces are of course not included, but we have confirmed that for finite systems the role of Coulomb is important~\cite{PhysRevC.90.027602}.  

In summary, we have extracted the free energy (density) for fermions and bosons in finite nuclei at subsaturation densities and finite temperatures using the Landau free energy technique. It was found that free-energy results for $\alpha$-particles are negative and very close to those of ideal Bose gases, whereas deuterons behave much like fermions. The $\alpha$-particle fraction was shown to be favored at all temperatures and densities explored in this work. The present results are consistent with the clusterization of nuclear matter into $\alpha$-particles. In the limit of zero temperature and ground-state density, the free energy discussed above reduces to the symmetry and pairing terms in the Weizs\"acker mass formula.

%\section{CONCLUSIONS}

\begin{acknowledgements}
This work was supported by the Robert A. Welch Foundation under Grant No. A-1266 and the U. S. Department of Energy under Grant No. DE-FG03-93ER-40773.
\end{acknowledgements}

%\bibliographystyle{alpha}
%\bibliographystyle{plain}
%\bibliographystyle{h-physrev3.bst}
%\bibliographystyle{apsrev}
%\bibliography{acompat,biblio}
\bibliography{Landau_FE_bosons_fermions}

\begin{thebibliography}{41}
\expandafter\ifx\csname natexlab\endcsname\relax\def\natexlab#1{#1}\fi
\expandafter\ifx\csname bibnamefont\endcsname\relax
  \def\bibnamefont#1{#1}\fi
\expandafter\ifx\csname bibfnamefont\endcsname\relax
  \def\bibfnamefont#1{#1}\fi
\expandafter\ifx\csname citenamefont\endcsname\relax
  \def\citenamefont#1{#1}\fi
\expandafter\ifx\csname url\endcsname\relax
  \def\url#1{\texttt{#1}}\fi
\expandafter\ifx\csname urlprefix\endcsname\relax\def\urlprefix{URL }\fi
\providecommand{\bibinfo}[2]{#2}
\providecommand{\eprint}[2][]{\url{#2}}

\bibitem[{\citenamefont{Scarpaci et~al.}(2010)}]{PhysRevC.82.031301}
\bibinfo{author}{\bibfnamefont{J.~A.} \bibnamefont{Scarpaci}}
  \bibnamefont{et~al.}, \bibinfo{journal}{Phys. Rev. C}
  \textbf{\bibinfo{volume}{82}}, \bibinfo{pages}{031301}
  (\bibinfo{year}{2010}).

\bibitem[{\citenamefont{Freer and Fynbo}(2014)}]{Freer20141}
\bibinfo{author}{\bibfnamefont{M.}~\bibnamefont{Freer}} \bibnamefont{and}
  \bibinfo{author}{\bibfnamefont{H.}~\bibnamefont{Fynbo}},
  \bibinfo{journal}{Prog. Part. Nucl. Phys.} \textbf{\bibinfo{volume}{78}},
  \bibinfo{pages}{1 } (\bibinfo{year}{2014}).

\bibitem[{\citenamefont{Kowalski et~al.}(2007)}]{PhysRevC.75.014601}
\bibinfo{author}{\bibfnamefont{S.}~\bibnamefont{Kowalski}}
  \bibnamefont{et~al.}, \bibinfo{journal}{Phys. Rev. C}
  \textbf{\bibinfo{volume}{75}}, \bibinfo{pages}{014601}
  (\bibinfo{year}{2007}).

\bibitem[{\citenamefont{Natowitz et~al.}(2010)}]{PhysRevLett.104.202501}
\bibinfo{author}{\bibfnamefont{J.~B.} \bibnamefont{Natowitz}}
  \bibnamefont{et~al.}, \bibinfo{journal}{Phys. Rev. Lett.}
  \textbf{\bibinfo{volume}{104}}, \bibinfo{pages}{202501}
  (\bibinfo{year}{2010}).

\bibitem[{\citenamefont{Wada et~al.}(2012)}]{PhysRevC.85.064618}
\bibinfo{author}{\bibfnamefont{R.}~\bibnamefont{Wada}} \bibnamefont{et~al.},
  \bibinfo{journal}{Phys. Rev. C} \textbf{\bibinfo{volume}{85}},
  \bibinfo{pages}{064618} (\bibinfo{year}{2012}).

\bibitem[{\citenamefont{Qin et~al.}(2012)}]{PhysRevLett.108.172701}
\bibinfo{author}{\bibfnamefont{L.}~\bibnamefont{Qin}} \bibnamefont{et~al.},
  \bibinfo{journal}{Phys. Rev. Lett.} \textbf{\bibinfo{volume}{108}},
  \bibinfo{pages}{172701} (\bibinfo{year}{2012}).

\bibitem[{\citenamefont{Horowitz and Schwenk}(2006)}]{Horowitz200655}
\bibinfo{author}{\bibfnamefont{C.}~\bibnamefont{Horowitz}} \bibnamefont{and}
  \bibinfo{author}{\bibfnamefont{A.}~\bibnamefont{Schwenk}},
  \bibinfo{journal}{Nucl. Phys.} \textbf{\bibinfo{volume}{A776}},
  \bibinfo{pages}{55 } (\bibinfo{year}{2006}).

\bibitem[{\citenamefont{Typel et~al.}(2014)\citenamefont{Typel, Wolter,
  R\"opke, and Blaschke}}]{Eur.Phys.J.A50.17}
\bibinfo{author}{\bibfnamefont{S.}~\bibnamefont{Typel}},
  \bibinfo{author}{\bibfnamefont{H.~H.} \bibnamefont{Wolter}},
  \bibinfo{author}{\bibfnamefont{G.}~\bibnamefont{R\"opke}}, \bibnamefont{and}
  \bibinfo{author}{\bibfnamefont{D.}~\bibnamefont{Blaschke}},
  \bibinfo{journal}{Eur. Phys. J.} \textbf{\bibinfo{volume}{A50}},
  \bibinfo{pages}{17} (\bibinfo{year}{2014}).

\bibitem[{\citenamefont{Danielewicz et~al.}(2002)\citenamefont{Danielewicz,
  Lacey, and Lynch}}]{Danielewicz22112002}
\bibinfo{author}{\bibfnamefont{P.}~\bibnamefont{Danielewicz}},
  \bibinfo{author}{\bibfnamefont{R.}~\bibnamefont{Lacey}}, \bibnamefont{and}
  \bibinfo{author}{\bibfnamefont{W.~G.} \bibnamefont{Lynch}},
  \bibinfo{journal}{Science} \textbf{\bibinfo{volume}{298}},
  \bibinfo{pages}{1592} (\bibinfo{year}{2002}).

\bibitem[{\citenamefont{Lattimer and Prakash}(2004)}]{Lattimer23042004}
\bibinfo{author}{\bibfnamefont{J.~M.} \bibnamefont{Lattimer}} \bibnamefont{and}
  \bibinfo{author}{\bibfnamefont{M.}~\bibnamefont{Prakash}},
  \bibinfo{journal}{Science} \textbf{\bibinfo{volume}{304}},
  \bibinfo{pages}{536} (\bibinfo{year}{2004}).

\bibitem[{\citenamefont{Li et~al.}(2008)\citenamefont{Li, Chen, and
  Ko}}]{li2008recent}
\bibinfo{author}{\bibfnamefont{B.-A.} \bibnamefont{Li}},
  \bibinfo{author}{\bibfnamefont{L.-W.} \bibnamefont{Chen}}, \bibnamefont{and}
  \bibinfo{author}{\bibfnamefont{C.~M.} \bibnamefont{Ko}},
  \bibinfo{journal}{Phys. Rep.} \textbf{\bibinfo{volume}{464}},
  \bibinfo{pages}{113} (\bibinfo{year}{2008}).

\bibitem[{\citenamefont{Giuliani et~al.}(2014)\citenamefont{Giuliani, Zheng,
  and Bonasera}}]{Giuliani2014116}
\bibinfo{author}{\bibfnamefont{G.}~\bibnamefont{Giuliani}},
  \bibinfo{author}{\bibfnamefont{H.}~\bibnamefont{Zheng}}, \bibnamefont{and}
  \bibinfo{author}{\bibfnamefont{A.}~\bibnamefont{Bonasera}},
  \bibinfo{journal}{Prog. Part. Nucl. Phys.} \textbf{\bibinfo{volume}{76}},
  \bibinfo{pages}{116 } (\bibinfo{year}{2014}).

\bibitem[{\citenamefont{Zheng and Bonasera}(2011)}]{Zheng:2010kg}
\bibinfo{author}{\bibfnamefont{H.}~\bibnamefont{Zheng}} \bibnamefont{and}
  \bibinfo{author}{\bibfnamefont{A.}~\bibnamefont{Bonasera}},
  \bibinfo{journal}{Phys. Lett.} \textbf{\bibinfo{volume}{B696}},
  \bibinfo{pages}{178} (\bibinfo{year}{2011}).

\bibitem[{\citenamefont{Zheng and Bonasera}(2012)}]{PhysRevC.86.027602}
\bibinfo{author}{\bibfnamefont{H.}~\bibnamefont{Zheng}} \bibnamefont{and}
  \bibinfo{author}{\bibfnamefont{A.}~\bibnamefont{Bonasera}},
  \bibinfo{journal}{Phys. Rev. C} \textbf{\bibinfo{volume}{86}},
  \bibinfo{pages}{027602} (\bibinfo{year}{2012}).

\bibitem[{\citenamefont{Zheng et~al.}(2012)\citenamefont{Zheng, Giuliani, and
  Bonasera}}]{Zheng201243}
\bibinfo{author}{\bibfnamefont{H.}~\bibnamefont{Zheng}},
  \bibinfo{author}{\bibfnamefont{G.}~\bibnamefont{Giuliani}}, \bibnamefont{and}
  \bibinfo{author}{\bibfnamefont{A.}~\bibnamefont{Bonasera}},
  \bibinfo{journal}{Nucl. Phys.} \textbf{\bibinfo{volume}{A892}},
  \bibinfo{pages}{43} (\bibinfo{year}{2012}).

\bibitem[{\citenamefont{Zheng et~al.}(2013)\citenamefont{Zheng, Giuliani, and
  Bonasera}}]{PhysRevC.88.024607}
\bibinfo{author}{\bibfnamefont{H.}~\bibnamefont{Zheng}},
  \bibinfo{author}{\bibfnamefont{G.}~\bibnamefont{Giuliani}}, \bibnamefont{and}
  \bibinfo{author}{\bibfnamefont{A.}~\bibnamefont{Bonasera}},
  \bibinfo{journal}{Phys. Rev. C} \textbf{\bibinfo{volume}{88}},
  \bibinfo{pages}{024607} (\bibinfo{year}{2013}).

\bibitem[{\citenamefont{Zheng et~al.}(2014)\citenamefont{Zheng, Giuliani, and
  Bonasera}}]{JPGNuclPartPhys.41.055109}
\bibinfo{author}{\bibfnamefont{H.}~\bibnamefont{Zheng}},
  \bibinfo{author}{\bibfnamefont{G.}~\bibnamefont{Giuliani}}, \bibnamefont{and}
  \bibinfo{author}{\bibfnamefont{A.}~\bibnamefont{Bonasera}},
  \bibinfo{journal}{J. Phys. G: Nucl. Part. Phys.}
  \textbf{\bibinfo{volume}{41}}, \bibinfo{pages}{055109}
  (\bibinfo{year}{2014}).

\bibitem[{\citenamefont{Wuenschel
  et~al.}(2009{\natexlab{a}})}]{Wuenschel2009578}
\bibinfo{author}{\bibfnamefont{S.}~\bibnamefont{Wuenschel}}
  \bibnamefont{et~al.}, \bibinfo{journal}{Nucl. Instrum. Methods A}
  \textbf{\bibinfo{volume}{604}}, \bibinfo{pages}{578}
  (\bibinfo{year}{2009}{\natexlab{a}}).

\bibitem[{\citenamefont{Schmitt et~al.}(1995)}]{Schmitt1995487}
\bibinfo{author}{\bibfnamefont{R.}~\bibnamefont{Schmitt}} \bibnamefont{et~al.},
  \bibinfo{journal}{Nucl. Instr. Meth. A} \textbf{\bibinfo{volume}{354}},
  \bibinfo{pages}{487 } (\bibinfo{year}{1995}).

\bibitem[{\citenamefont{Kohley}(2010)}]{KohleyPhD}
\bibinfo{author}{\bibfnamefont{Z.}~\bibnamefont{Kohley}}, Ph.D. thesis,
  \bibinfo{school}{Texas A$\&$M University} (\bibinfo{year}{2010}).

\bibitem[{\citenamefont{Kohley et~al.}(2011)}]{PhysRevC.83.044601}
\bibinfo{author}{\bibfnamefont{Z.}~\bibnamefont{Kohley}} \bibnamefont{et~al.},
  \bibinfo{journal}{Phys. Rev. C} \textbf{\bibinfo{volume}{83}},
  \bibinfo{pages}{044601} (\bibinfo{year}{2011}).

\bibitem[{\citenamefont{Kohley et~al.}(2012)}]{PhysRevC.86.044605}
\bibinfo{author}{\bibfnamefont{Z.}~\bibnamefont{Kohley}} \bibnamefont{et~al.},
  \bibinfo{journal}{Phys. Rev. C} \textbf{\bibinfo{volume}{86}},
  \bibinfo{pages}{044605} (\bibinfo{year}{2012}).

\bibitem[{\citenamefont{Wuenschel et~al.}(2010)}]{Wuenschel20101}
\bibinfo{author}{\bibfnamefont{S.}~\bibnamefont{Wuenschel}}
  \bibnamefont{et~al.}, \bibinfo{journal}{Nucl. Phys.}
  \textbf{\bibinfo{volume}{A843}}, \bibinfo{pages}{1} (\bibinfo{year}{2010}).

\bibitem[{\citenamefont{Marini et~al.}(2013)}]{Marini201380}
\bibinfo{author}{\bibfnamefont{P.}~\bibnamefont{Marini}} \bibnamefont{et~al.},
  \bibinfo{journal}{Nucl. Instr. Meth. A} \textbf{\bibinfo{volume}{707}},
  \bibinfo{pages}{80 } (\bibinfo{year}{2013}).

\bibitem[{\citenamefont{Wuenschel
  et~al.}(2009{\natexlab{b}})}]{PhysRevC.79.061602}
\bibinfo{author}{\bibfnamefont{S.}~\bibnamefont{Wuenschel}}
  \bibnamefont{et~al.}, \bibinfo{journal}{Phys. Rev. C}
  \textbf{\bibinfo{volume}{79}}, \bibinfo{pages}{061602}
  (\bibinfo{year}{2009}{\natexlab{b}}).

\bibitem[{\citenamefont{Bonasera et~al.}(2008)}]{PhysRevLett.101.122702}
\bibinfo{author}{\bibfnamefont{A.}~\bibnamefont{Bonasera}}
  \bibnamefont{et~al.}, \bibinfo{journal}{Phys. Rev. Lett.}
  \textbf{\bibinfo{volume}{101}}, \bibinfo{pages}{122702}
  (\bibinfo{year}{2008}).

\bibitem[{\citenamefont{Huang et~al.}(2010)}]{PhysRevC.81.044618}
\bibinfo{author}{\bibfnamefont{M.}~\bibnamefont{Huang}} \bibnamefont{et~al.},
  \bibinfo{journal}{Phys. Rev. C} \textbf{\bibinfo{volume}{81}},
  \bibinfo{pages}{044618} (\bibinfo{year}{2010}).

\bibitem[{\citenamefont{Tripathi et~al.}(2011)}]{PhysRevC.83.054609}
\bibinfo{author}{\bibfnamefont{R.}~\bibnamefont{Tripathi}}
  \bibnamefont{et~al.}, \bibinfo{journal}{Phys. Rev. C}
  \textbf{\bibinfo{volume}{83}}, \bibinfo{pages}{054609}
  (\bibinfo{year}{2011}).

\bibitem[{\citenamefont{Mabiala et~al.}(2013)}]{PhysRevC.87.017603}
\bibinfo{author}{\bibfnamefont{J.}~\bibnamefont{Mabiala}} \bibnamefont{et~al.},
  \bibinfo{journal}{Phys. Rev. C} \textbf{\bibinfo{volume}{87}},
  \bibinfo{pages}{017603} (\bibinfo{year}{2013}).

\bibitem[{\citenamefont{Fisher}(1967)}]{Fisher1967}
\bibinfo{author}{\bibfnamefont{M.~E.} \bibnamefont{Fisher}},
  \bibinfo{journal}{Rep. Prog. Phys.} \textbf{\bibinfo{volume}{30}},
  \bibinfo{pages}{615} (\bibinfo{year}{1967}).

\bibitem[{\citenamefont{Minich et~al.}(1982)}]{Minich1982458}
\bibinfo{author}{\bibfnamefont{R.}~\bibnamefont{Minich}} \bibnamefont{et~al.},
  \bibinfo{journal}{Phys. Lett.} \textbf{\bibinfo{volume}{B118}},
  \bibinfo{pages}{458 } (\bibinfo{year}{1982}).

\bibitem[{\citenamefont{Bonasera et~al.}(2000)}]{Bonasera20001}
\bibinfo{author}{\bibfnamefont{A.}~\bibnamefont{Bonasera}}
  \bibnamefont{et~al.}, \bibinfo{journal}{La Rivista Del Nuovo Cimento}
  \textbf{\bibinfo{volume}{23}}, \bibinfo{pages}{1} (\bibinfo{year}{2000}).

\bibitem[{\citenamefont{Tripathi et~al.}(2012)}]{S021830131250019X}
\bibinfo{author}{\bibfnamefont{R.}~\bibnamefont{Tripathi}}
  \bibnamefont{et~al.}, \bibinfo{journal}{Int. J. Mod. Phys. E}
  \textbf{\bibinfo{volume}{21}}, \bibinfo{pages}{1250019}
  (\bibinfo{year}{2012}).

\bibitem[{\citenamefont{Huang}(1987)}]{K.Huang}
\bibinfo{author}{\bibfnamefont{K.}~\bibnamefont{Huang}},
  \emph{\bibinfo{title}{Statistical Mechanics}} (\bibinfo{publisher}{Wiley $\&$
  Sons, New York}, \bibinfo{year}{1987}).

\bibitem[{\citenamefont{Hagel et~al.}(2012)}]{PhysRevLett.108.062702}
\bibinfo{author}{\bibfnamefont{K.}~\bibnamefont{Hagel}} \bibnamefont{et~al.},
  \bibinfo{journal}{Phys. Rev. Lett.} \textbf{\bibinfo{volume}{108}},
  \bibinfo{pages}{062702} (\bibinfo{year}{2012}).

\bibitem[{\citenamefont{Mabiala et~al.}(2014)}]{PhysRevC.90.027602}
\bibinfo{author}{\bibfnamefont{J.}~\bibnamefont{Mabiala}} \bibnamefont{et~al.},
  \bibinfo{journal}{Phys. Rev. C} \textbf{\bibinfo{volume}{90}},
  \bibinfo{pages}{027602} (\bibinfo{year}{2014}).

\bibitem[{\citenamefont{Mabiala et~al.}(2015)}]{PhysRevC.92.024605}
\bibinfo{author}{\bibfnamefont{J.}~\bibnamefont{Mabiala}} \bibnamefont{et~al.},
  \bibinfo{journal}{Phys. Rev. C} \textbf{\bibinfo{volume}{92}},
  \bibinfo{pages}{024605} (\bibinfo{year}{2015}).

\bibitem[{\citenamefont{Marini et~al.}(2016)}]{Marini2016194}
\bibinfo{author}{\bibfnamefont{P.}~\bibnamefont{Marini}} \bibnamefont{et~al.},
  \bibinfo{journal}{Phys. Lett.} \textbf{\bibinfo{volume}{B756}},
  \bibinfo{pages}{194 } (\bibinfo{year}{2016}).

\bibitem[{\citenamefont{Typel et~al.}(2010)}]{PhysRevC.81.015803}
\bibinfo{author}{\bibfnamefont{S.}~\bibnamefont{Typel}} \bibnamefont{et~al.},
  \bibinfo{journal}{Phys. Rev. C} \textbf{\bibinfo{volume}{81}},
  \bibinfo{pages}{015803} (\bibinfo{year}{2010}).

\bibitem[{\citenamefont{Landau and Lifshitz}(1980)}]{Laundau.Lifshitz}
\bibinfo{author}{\bibfnamefont{L.~D.} \bibnamefont{Landau}} \bibnamefont{and}
  \bibinfo{author}{\bibfnamefont{E.~M.} \bibnamefont{Lifshitz}},
  \emph{\bibinfo{title}{Statistical Physics}} (\bibinfo{publisher}{Pergamon,
  New York}, \bibinfo{year}{1980}).

\bibitem[{\citenamefont{Schreck et~al.}(2001)}]{PhysRevLett.87.080403}
\bibinfo{author}{\bibfnamefont{F.}~\bibnamefont{Schreck}} \bibnamefont{et~al.},
  \bibinfo{journal}{Phys. Rev. Lett.} \textbf{\bibinfo{volume}{87}},
  \bibinfo{pages}{080403} (\bibinfo{year}{2001}).

\end{thebibliography}

\end{document}